# Effect of Vacancies on Hydrogen Mobility and Trapping in Elemental Fe and Cr: A DFT and kMC Study

**Vallinathan K[1,2], Gurpreet Kaur[1,2], Sharat Chandra[1,2]**

**[1]Material Science Group, Indira Gandhi Centre for Atomic Research, Kalpakkam 603102, Tamil Nadu**

**[2]HBNI, Anushakti Nagar, Mumbai 400094, India**

**Email:** vallinathan8@igcar.gov.in, gurpreet@igcar.gov.in , sharat@igcar.gov.in

**Abstract:** Hydrogen–vacancy interactions play an important role in governing hydrogen transport and embrittlement in body-centered cubic (BCC) metals. In this study, a multiscale approach combining density functional theory (DFT) and kinetic Monte Carlo (kMC) simulations is employed to investigate hydrogen behavior in BCC Fe and Cr. The DFT-calculated binding energies and Bader charge analysis indicate stronger hydrogen trapping in Cr than in Fe. Migration and detrapping energy barriers are determined using the climbing-image nudged elastic band method, showing that the detrapping energy generally decreases with increasing hydrogen occupancy. However, the sixth hydrogen atom in Fe exhibits a finite barrier, contrary to some previous reports. kMC simulations are then used to evaluate hydrogen diffusion over extended time and length scales. The results demonstrate that vacancy defects significantly reduce hydrogen mobility and increase the effective activation energy, with a more pronounced effect observed in Cr due to stronger trapping. The combined DFT–kMC framework provides detailed insight into the mechanisms of hydrogen trapping, detrapping, and diffusion in BCC metals, offering important implications for understanding hydrogen embrittlement in structural materials.

## 1. Introduction

The presence of hydrogen (H) in metallic alloys degrades their mechanical properties, reducing toughness, ductility, and fatigue resistance—a phenomenon known as hydrogen embrittlement (HE) [1-2]. HE is broadly classified into two types: internal hydrogen embrittlement (IHE) and environmental hydrogen embrittlement (EHE) [3]. IHE arises from pre-existing hydrogen within the material and is therefore limited by the available hydrogen supply, whereas EHE occurs when a material is subjected to mechanical loading while simultaneously exposed to a hydrogen environment. Several mechanisms have been proposed to explain HE, including hydride formation [4-6], hydrogen-enhanced decohesion (HEDE) [7-9], hydrogen-enhanced local plasticity (HELP) [10-12], and hydrogen-enhanced strain-induced vacancies (HESIV) [13]. Iron (Fe) and chromium (Cr) are the primary constituents of steel and stainless steel alloys, which are widely used in medical, aerospace, automotive, and nuclear reactor applications. Hydrogen isotopes such as deuterium and tritium are promising fuels for fusion reactors [14]; however, their trapping within structural materials poses a significant challenge [15]. Understanding hydrogen diffusion in these materials is critical, as hydrogen's high mobility allows it to accumulate at microstructural defects, such as crack tips, grain boundaries, and vacancies [16-18]. In this work, we investigate the effect of vacancies on hydrogen mobility in Fe and Cr systems. Atomistic simulations of hydrogen diffusivity have been extensively studied in pristine Fe [7], but remain comparatively limited in Cr [19]; moreover, the role of vacancies in influencing hydrogen diffusion has received relatively little attention. In this study, first-principles density functional theory (DFT) calculations are employed to determine defect formation and migration energies, while the kinetic Monte Carlo (kMC) method is used to examine hydrogen mobility and trapping behavior. Although hydrogen diffusion in pristine Fe, Cr, and their alloys [20] has been studied, investigations addressing diffusion in the presence of vacancies are relatively limited. Some recent kMC studies have examined hydrogen diffusion in defective systems such as Fe and W, and in V systems [21-22]. A. Khosravi et al. analyzed hydrogen trapping and detrapping energies from a monovacancy using the k-ARTn method with an embedded atom method (EAM) potential, explaining the observed detrapping energy trends. The investigation of hydrogen diffusion in pristine Cr remains limited, and is even more sparse in the presence of defects. Miedie et al. [23] studied hydrogen diffusion in ferritic steels and demonstrated the influence of vacancies and Cr on diffusion behavior. In the present work, we calculate all six hydrogen detrapping energies from a monovacancy, the binding energy with a monovacancy, and the variation in vacancy formation energy with increasing hydrogen concentration near the vacancy using first-principles calculations. Furthermore, kMC simulations are used to capture hydrogen trapping at monovacancies, evaluate diffusion coefficients in the presence of vacancies, and analyze changes in activation energy with increasing vacancy concentration.

## 2. Methodology

### a. DFT Calculation

Density functional theory (DFT) calculations were carried out using the Vienna *Ab initio* Simulation Package (VASP)[24], [25], [26]. The exchange–correlation interactions were described within the generalized gradient approximation (GGA) using the Perdew–Burke–Ernzerhof (PBE) [27] functional. Electron–ion interactions were treated using the projector augmented-wave (PAW) [28] method, and a plane-wave kinetic energy cutoff of 500 eV was employed. Supercells containing 54 atoms were constructed for body-centered cubic (BCC) Fe and Cr, and the Brillouin-zone integrations were performed using Γ-centered Monkhorst–Pack [29] k-point meshes of 5 × 5 × 5. Electronic self-consistency was achieved with an energy convergence criterion of $10^{-7}$ eV, and atomic structures were fully relaxed until the residual forces on each atom were less than 0.01 eV/Å. Hydrogen migration energy barriers were determined using the climbing nudged elastic band (ClNEB) [30] method, allowing full relaxation of all atoms until the force on each atom converged to 0.01 eV/Å. Zero-point energy (ZPE) corrections were evaluated by constructing the Hessian matrix for the hydrogen atom, from which the vibrational normal modes and corresponding frequencies were obtained. The ZPE was calculated using

$$ZPE = \frac{1}{2}\sum_i h\nu_i \tag{1.0}$$

where $h$ is Planck's constant and $\nu$ denotes the vibrational frequencies of the hydrogen atom.

### b. kMC Model

The kinetic Monte Carlo (kMC) method is a stochastic simulation technique widely employed to investigate the time evolution of systems governed by thermally activated processes. In the present study, hydrogen atoms were initially distributed randomly over the tetrahedral (tetra) sites in the BCC lattices of Fe and Cr. The event generation of this kMC is based on transition state theory (TST). The jump rate ($\Gamma_{i\to j}$) for a hydrogen atom migrating from site $i$ to site $j$ is expressed as

$$\Gamma_{i\to j} = \nu_0 \exp\left(\frac{-E_m}{k_B T}\right) \tag{1.1}$$

where $\nu_0$ is the attempt frequency, taken as $1\times 10^{13}$ $s^{-1}$, $E_m$ is the migration energy barrier, $k_B$ is the Boltzmann constant, and $T$ is the absolute temperature. The total transition rate in the system is given by

$$\Gamma_{tot} = \sum_{\boldsymbol{i,j}} \Gamma_{i\to j} \tag{1.2}$$

The time increment associated with each kMC step is determined using

$$\Delta t = -\frac{\ln(r)}{\Gamma_{tot}} \tag{1.3}$$

where $r$ is a uniformly distributed random number between 0 and 1. The kMC simulation proceeds until the prescribed total simulation time is reached, after which the hydrogen diffusion coefficient is evaluated using the Einstein relation,

$$D = \frac{\langle |\boldsymbol{r}_i(t) - \boldsymbol{r}_i(0)| \rangle^2}{2dt} \tag{1.4}$$

where $\boldsymbol{r}_i(t)$ and , $\boldsymbol{r}_i(0)$ denote the positions of the hydrogen atom at time $t$ and at the initial time, respectively. The parameter $d$ represents the dimensionality of the system, which is set to 3 in this work. To investigate hydrogen diffusion in the presence of vacancies, a vacancy was randomly introduced at a lattice site in Fe and Cr. A trapping radius was defined around the vacancy; when a hydrogen atom reaches the boundary of the radius using the trapping migration energy to enters into the trapping region, then it must overcome the detrapping migration energy to escape from the vacancy. A trapping radius of 2 Å was employed for all systems. Owing to the highly repulsive interaction between hydrogen atoms [31], H–H interactions were not included in the kMC model. Vacancy migration was also neglected because its migration energy barrier is significantly higher [32]than that of hydrogen. The maximum number of hydrogen atoms allowed to be trapped at a single vacancy was set to six, based on reported literature values for Fe and Cr [33]. To reduce statistical noise, each kMC simulation was repeated eighteen times at each temperature. The reported diffusion coefficients and total number of trapped hydrogen atoms were obtained by averaging over all simulation runs.

## 3. Results and Discussion

### a. Vacancy Formation Energy

In body-centred cubic (BCC), the two primary interstitial positions are the tetrahedral (tetra) site and the octahedral (octa) site. For hydrogen, the solute formation energy $E_{sol}^f$ was calculated in Fe and Cr to determine the preferred interstitial site. The solute formation energy $E_{sol}^f$ is defined as

$$E_{sol}^f = E_H - E_{perf} - \frac{1}{2} E_{H_2} \tag{1.5}$$

where $E_H$ is the total energy of the system containing a hydrogen solute, $E_{perf}$ is the total energy of the corresponding perfect system without hydrogen, and $E_{H_2}$ is the reference energy of an isolated $H_2$ molecule. For both Fe and Cr, hydrogen preferentially occupies the tetrahedral site rather than the octahedral site because H prefers to reside in interstitial sites with high charge density [34]. The calculated formation energy of H in Fe is 0.20 eV, which is in good agreement with the literature [7], while the calculated formation energy in Cr is 0.46 eV. Even when hydrogen was initially placed at an octahedral site, structural relaxation drove the atom toward a neighbouring tetrahedral site, in good agreement with previous studies [19][35]. The vacancy formation energies of Fe and Cr were calculated in the presence of hydrogen. Hydrogen was initially placed at the tetra site near the vacancy in both Fe and Cr. After structural relaxation, the hydrogen atom moved towards the vacancy and was found near the octahedral site, located 0.23 Å and 0.34 Å away from the ideal octahedral position in Fe and Cr, respectively. These observations are in very good agreement with previous works [34, 36]. As hydrogen was progressively introduced into the system, it was found that up to six hydrogen atoms could occupy the vicinity of a single vacancy in both Fe and Cr. The zero-point energy (ZPE) contribution of hydrogen was found to have a negligible effect on the vacancy formation and binding energies [37, 38]; therefore, ZPE corrections were not included in the present calculations. The vacancy formation energy of a pure system without hydrogen is defined as

$$E_{vac}^f = E_{def} - \frac{N-1}{N} E_{perf} \tag{1.6}$$

where $E_{def}$ is the total energy of a supercell containing N−1 atoms and one vacancy and $E_{perf}$ is the total energy of a perfect supercell containing N atoms. The vacancy formation energy of a system containing hydrogen in the vicinity of a vacancy is defined as

$$E^{f}_{vac,H} = E_{VH_n} - \frac{N-1}{N} E_{perf} - n\mu_H \quad (1.7)$$

where $E_{VH_n}$ is the total energy of a supercell containing a monovacancy and n hydrogen atoms trapped at the vacancy, $E_{perf}$ is the energy of the defect-free supercell, and $\mu_H$ is the reference energy of an isolated $H_2$ molecule, typically defined as $\frac{1}{2} E_{H_2}$. The introduction of hydrogen in the vicinity of a vacancy leads to a gradual decrease in the vacancy formation energy. Figure. 1 illustrates the variation of vacancy formation energy with the addition of H in proximity to the vacancy.

**b. Binding energy of H with a monovacancy**

The binding energy of hydrogen to a monovacancy in Fe and Cr was calculated. Previous studies have shown that up to six hydrogen atoms can occupy the vicinity of a single vacancy in Fe and Cr. The binding energy between hydrogen atoms and a vacancy is defined as

$$E_{bin} = E_{vac} - E_{vac+mH} + m\,(E_H - E_{pure}) \quad (1.8)$$

where $E_{vac}$ is the total energy of the supercell containing a vacancy, $E_{vac+mH}$ is the total energy of the supercell containing a vacancy and *m* hydrogen atoms, $E_H$ is the total energy of a supercell containing a hydrogen atom at the energetically favourable interstitial site (tetra site or octa site), $E_{pure}$ is the total energy of the perfect (defect-free) supercell, and *m* is the number of hydrogen atoms bound to the vacancy. Initially, H was placed near the vacancy in both the Fe and Cr systems. After structural relaxation, H moved toward the vacancy center in both cases, however, it moved closer to the vacancy center in the Cr system (1.083 Å) than in the Fe system (1.185 Å). The solute formation energy trend of H in Fe and Cr shows that the formation energy of H is higher in the Cr system than in the Fe system, indicating that H is less stable at the interstitial site in Cr than in Fe. As a result, H moves more strongly toward the vacancy in Cr than in Fe. This behavior is further supported by QTAIM analysis, which shows that the bond critical point (BCP) electron density is higher for H in Fe than in Cr. Consistently, the FeH bond length in the Fe vacancy case (1.691 Å) is shorter than the CrH bond length in the Cr vacancy case (1.762 Å). The Bader charge analysis shows that the total charge gained by H is 0.30 e in Fe and 0.35 e in Cr, with the Cr value consistent with the recent literature [39]. As shown in Figure. 1, Cr exhibits the higher hydrogen vacancy binding energy, than by Fe, indicating a stronger tendency for hydrogen trapping at vacancies in Cr compared to Fe.

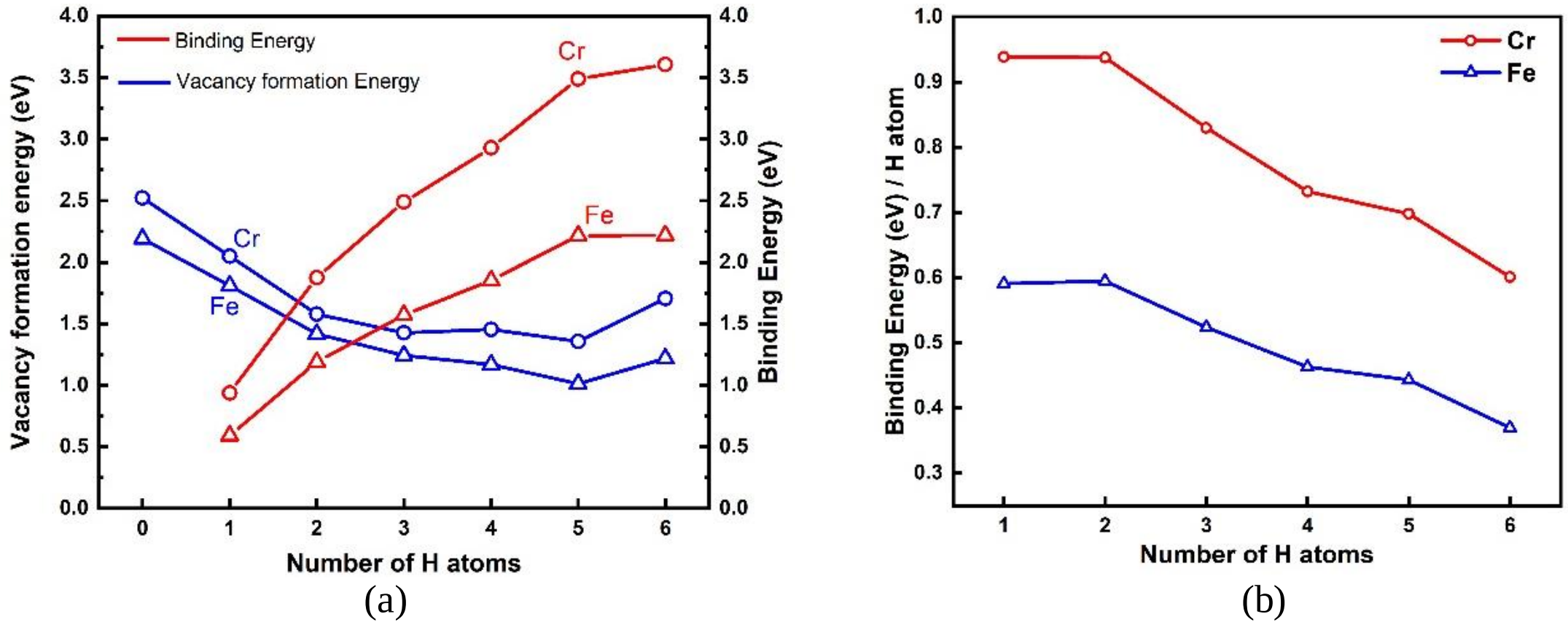

Figure 1, (a) represents the comparison of binding energy and vacancy formation energy with the addition of hydrogen near the vacancy. (b) represents the binding energy per hydrogen atom.

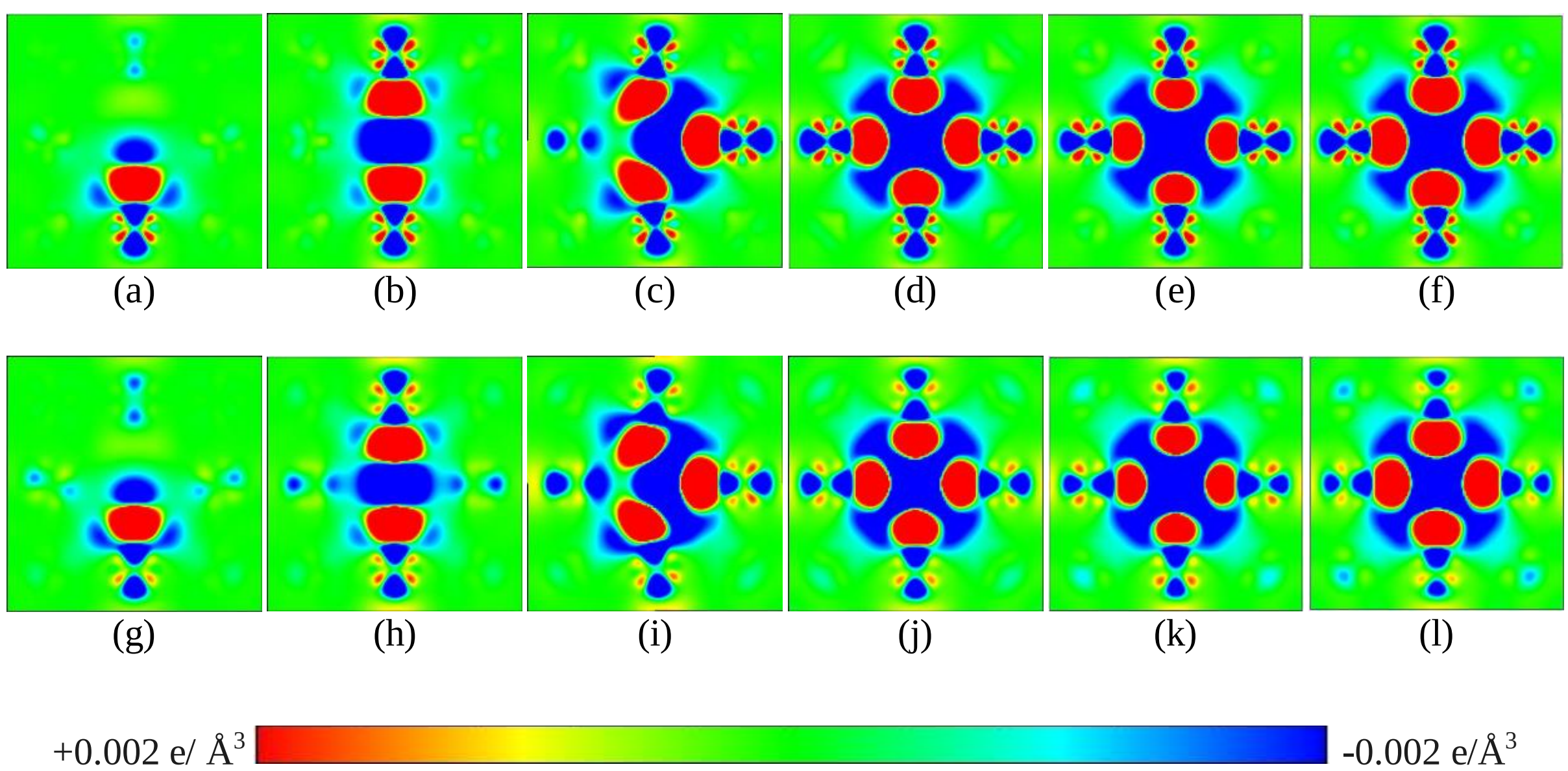

Figure 2, (a) to (f) illustrate the charge density difference for the Fe system containing a vacancy surrounded by 1 to 6 hydrogen atoms, respectively. (g) to (l) show the corresponding charge density difference for the Cr system containing a vacancy with 1 to 6 hydrogen atoms, respectively.

Table 1. Binding energy, binding energy per H atom, vacancy formation energy, and Bader charge of the H atom with the addition of H atoms to the monovacancy in the Fe and Cr systems.

| | Number of H atoms | Binding Energy (eV) | Binding energy per H (eV) | Vacancy formation energy (eV) | Bader Charge of H atoms |
|---|---|---|---|---|---|
| | 0 | | | 2.522 | |
| | 1 | 0.93872 | 0.93872 | 2.0488 | 1.35 |
| | 2 | 1.87508 | 0.93754 | 1.5773 | 1.36,1.36 |
| Cr | 3 | 2.48956 | 0.8298 | 1.4276 | 1.36,1.36,1.39 |
| | 4 | 2.92869 | 0.7321 | 1.4534 | 1.35,1.37,1.36,1.36 |
| | 5 | 3.49003 | 0.698 | 1.3568 | 1.35,1.35,1.35,1.35,1.40 |
| | 6 | 3.60637 | 0.601 | 1.7053 | 1.36,1.38,1.37,1.37,1.39,1.36 |
| | 0 | | | 2.191 | |
| | 1 | 0.59082 | 0.59082 | 1.8083 | 1.30 |
| | 2 | 1.18914 | 0.59455 | 1.4176 | 1.30,1.30 |
| Fe | 3 | 1.57115 | 0.5237 | 1.2431 | 1.30,1.30,1.34 |
| | 4 | 1.85261 | 0.4631 | 1.1692 | 1.32,1.32,1.32,1.32 |
| | 5 | 2.2168 | 0.4432 | 1.0125 | 1.27,1.27,1.27,1.27,1.33 |
| | 6 | 2.21933 | 0.3698 | 1.2176 | 1.30,1.30,1.30,1.30,1.30,1.30 |

### c. Migration and Detrapping energy of H

The mobility of H in both Fe and Cr occurs via tetra to tetra site migration. The hydrogen migration energy, including ZPE corrections, was calculated for pristine Fe and Cr. The results indicate that H diffusion is slightly faster in Fe than in Cr, consistent with the lower migration energy in Fe. The calculated migration energies are 0.05 eV for Fe and 0.059 eV for Cr, in good agreement with previous reports [7][40]. Due to the low migration barriers, H atoms have a high probability of encountering vacancies. Once trapped at a vacancy, the migration pathways and associated energy barriers are significantly altered, and the diffusion mechanism changes from tetra–tetra migration to a path involving octa sites and second-nearest-neighbour tetra sites. As discussed by K. Oshawa et al. [33], the equilibrium configuration of hydrogen becomes increasingly complex when four or more H atoms occupy a vacancy. In such cases, multiple metastable states and degenerate ground states are observed in both Fe and Cr. In the present work, the detrapping energies of all six H atoms from the vacancy are calculated for both Fe and Cr. The detrapping energy of H generally decreases with increasing H occupancy at the vacancy. However, the detrapping energy of the fourth H atom is lower than that of the fifth H atom in both Fe and Cr. This behavior can be explained by the atomic arrangement, the first four H atoms are symmetrically distributed in the plane of the octa site, whereas the fifth H atom occupies a position above this plane. As a result, the fifth H atom experiences reduced H–H repulsion compared to the fourth H atom. The migration energy between tetra sites and the detrapping energy from the monovacancy in both Fe and Cr are shown in Figure 3. Cl-NEB calculations indicate that the chosen initial states are metastable for both Fe and Cr. For the sixth H atom in Fe, the first intermediate image has a slightly lower energy than the initial state, confirming its metastable nature. In Cr, for the fourth H atom, both the first and second images are nearly degenerate and lie approximately 0.0094 eV below the initial state. Similarly, for the sixth H atom in Cr, the first and second images are about 0.01 eV lower in energy than the initial configuration. These results indicate the presence of lower-energy configurations along the migration pathway. A previous first-principles study [33] showed that zero-point energy (ZPE) modifies absolute binding energies but does not significantly affect the qualitative trends of H–vacancy configurations in BCC metals. Therefore, ZPE corrections were not included in the calculation of H detrapping energies. The present results for H detrapping from vacancies in Fe are in good agreement with those reported by A. Khosravi et al. [21].The author claims that the sixth H atom undergoes barrierless detrapping and that the detrapping energy decreases progressively with increasing H content; however, this trend is not observed in the present study. In our results, the sixth H atom in Fe exhibits a finite detrapping energy, while the reverse jump to the vacancy (i.e., the trapping energy) is higher than the corresponding detrapping energy. Table 2 compares the detrapping energies of H from vacancies in Fe and Cr with those reported in previous studies on the Fe system.

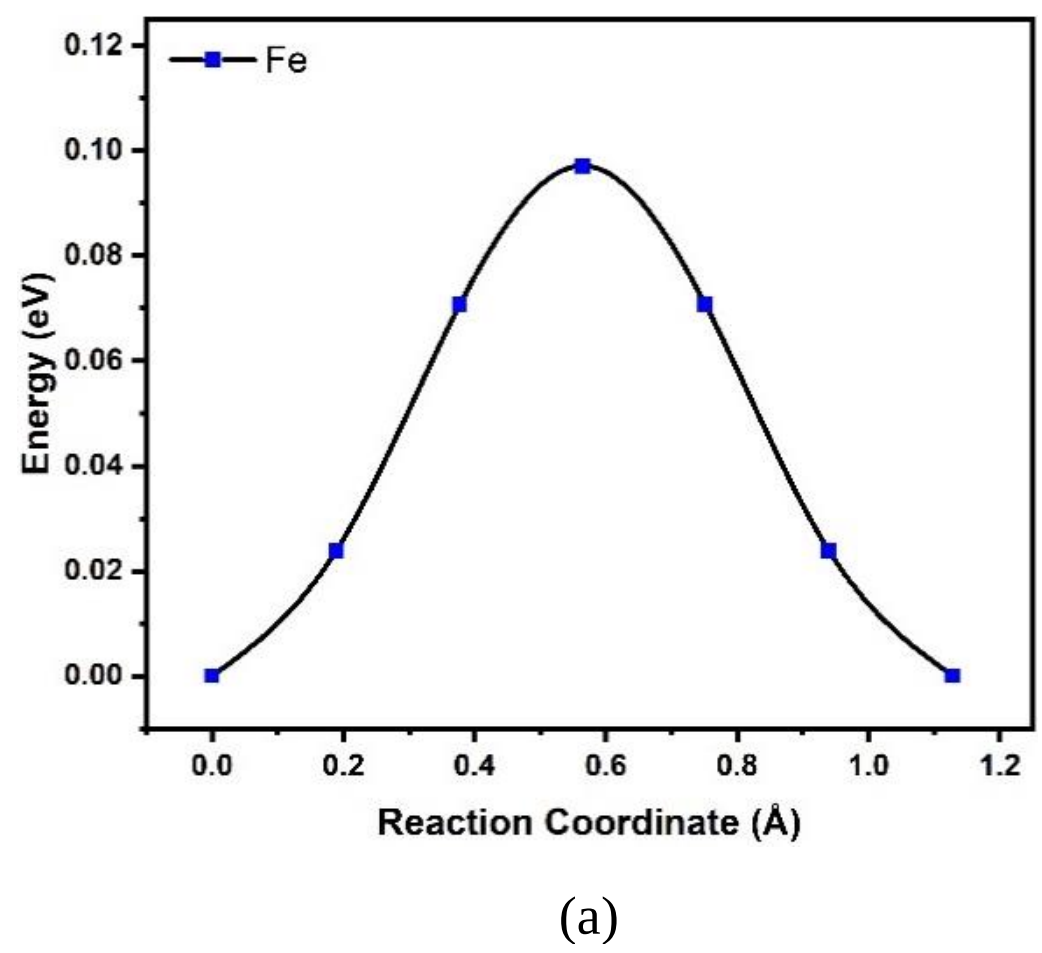

(a)

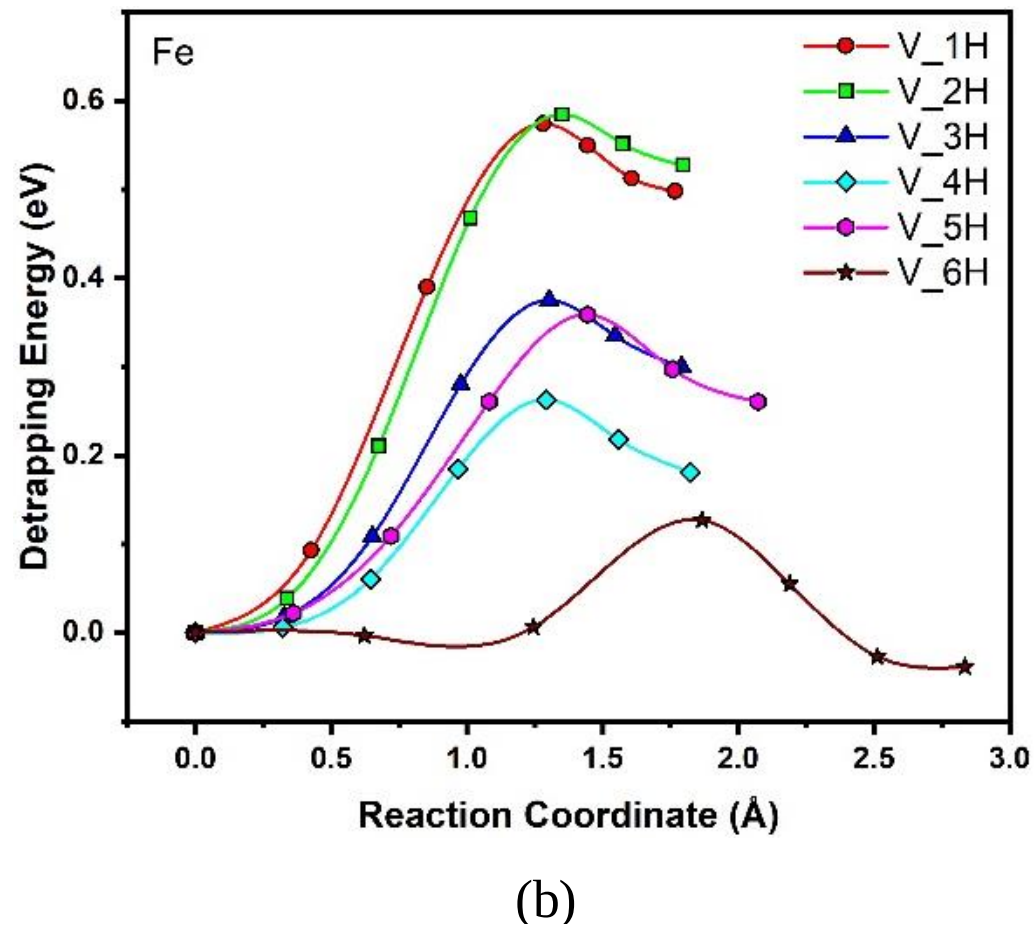

(b)

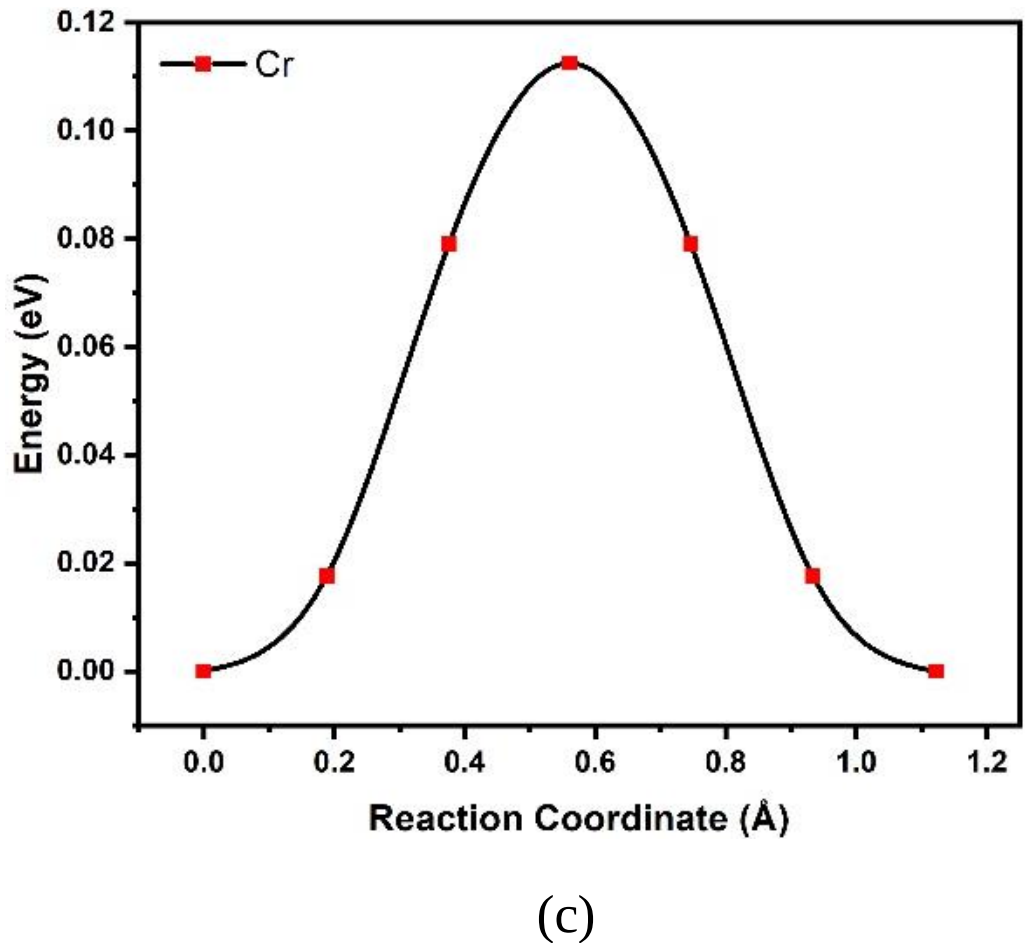

(c)

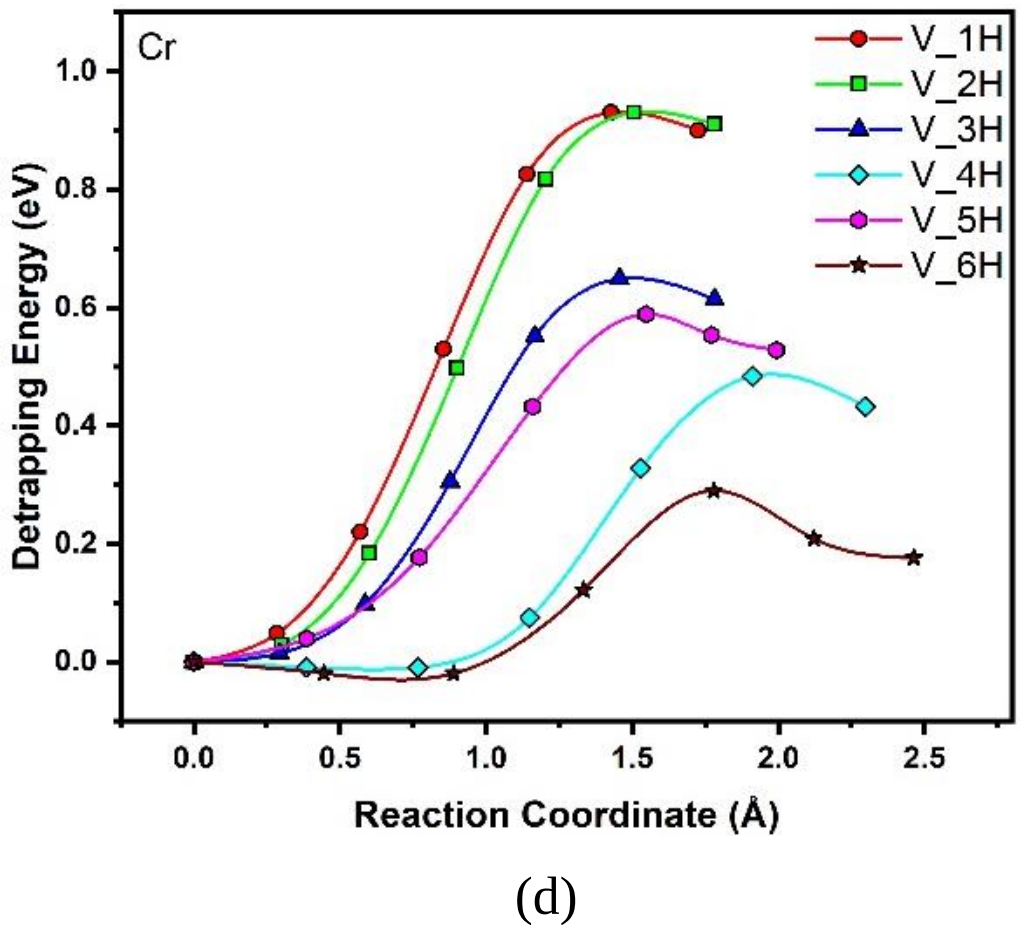

(d)

Figure 3. (a) Migration energy barrier for hydrogen between tetra sites in Fe. (b) Detrapping energy barrier for hydrogen migration from an Fe monovacancy. (c) Migration energy barrier for hydrogen between tetra sites in Cr. (d) Detrapping energy barrier for hydrogen migration from a Cr monovacancy.

Table 2 presents a comparison of the detrapping energies of H in Fe and Cr with available literature data. The results of the present work show trends that differ from those reported previously [21]. In particular, earlier study suggest that the detrapping energy of the fifth H atom is comparable to the migration energy of H in the absence of vacancies, while the sixth H atom is reported to undergo barrierless detrapping. This may be because of the EAM potential used to calculate the migration energy barrier.

| Number of H atoms | Fe (present work) (eV) | Fe [21] (eV) | Cr(present work) (eV) |
|---|---|---|---|
| 1 | 0.574 | 0.540 | 0.930 |
| 2 | 0.584 | 0.420 | 0.930 |
| 3 | 0.375 | 0.340 | 0.648 |
| 4 | 0.262 | 0.203 | 0.483 |
| 5 | 0.358 | 0.046 | 0.588 |
| 6 | 0.126 | Barrierless | 0.290 |

### d. Diffusion Coefficient of H in Fe and Cr

The diffusion coefficient of H in both pristine and vacancy containing systems was investigated using kinetic Monte Carlo (kMC) simulations. A 50 × 50 × 50 supercell was employed for both Fe and Cr, with 0.04% H (100 H) atoms randomly distributed at tetra interstitial sites. To examine the effect of vacancies on hydrogen diffusion, three vacancy concentrations were considered 40 ppm (10vac), 60 ppm (15vac), and 80 ppm (20vac). The simulation time in the kMC calculations varies with temperature. At lower temperatures (e.g., 300 K), the number of diffusion events is limited, and the simulation is extended up to $1 \times 10^{-6}$ s. At higher temperatures (500 K, 700 K, 900 K, and 1100 K), the simulation times are $5 \times 10^{-7}$ s, $1 \times 10^{-7}$ s, $5 \times 10^{-8}$ s, and $1 \times 10^{-8}$ s, respectively. The presence of vacancies reduces the mobility of H atoms. This effect is more pronounced in Cr than in Fe. Correspondingly, the activation energy for H migration increases with increasing vacancy concentration, as shown in Figure 4(c) and 4(d). The trend observed in Fe is in good agreement with previous studies [41]. The hydrogen diffusion coefficient decreases systematically with increasing vacancy concentration. As shown in Figure 4 (a), at 300K in Fe, the H in the pristine system exhibits a diffusion coefficient of $8.059\times10^{-9}$ m²/s. With the introduction of 10vac, the value decreases to $4.940\times10^{-9}$ m²/s, corresponding to a reduction of approximately 38.7% relative to the pristine system. Further increasing to 15vac yields a diffusion coefficient of $3.844\times10^{-9}$ m²/s, representing an overall reduction of approximately 52.3% from the pristine value. At 20vac, the diffusion coefficient further decreases to 2.593×10$^{-9}$ m²/s, corresponding to a total reduction of approximately 67.8% compared to the pristine system. These results clearly indicate a monotonic suppression of hydrogen diffusion with increasing vacancy concentration. Similarly, as shown in Figure 4 (b), at 300 K, the pristine Cr system exhibits a diffusion coefficient of 5.192×10$^{-9}$ m²/s. At 10vac, the diffusion coefficient decreases to 2.504×10$^{-9}$ m²/s, corresponding to a reduction of approximately 51.8% relative to the pristine system. For 15vac, the diffusion coefficient further decreases to 1.238×10$^{-9}$ m²/s, representing a reduction of approximately 76.1% relative to the pristine system. At 20vac, the diffusion coefficient drops to 1.526×10$^{-9}$ m²/s, corresponding to an overall reduction of approximately 97.1% relative to the pristine system. These results indicate that in Cr, vacancies act as stronger trapping sites for H atoms, although the trapping effect decreases with increasing temperature. The kMC results provide insight into the trapping behavior of H at vacancies. This observation aligns well with the total trapped H at 300 K, where the number of trapped H atoms is higher at 20vac in both Fe and Cr. Vacancy in Cr traps a larger number of H atoms, as shown in Figure 4 (f).

.

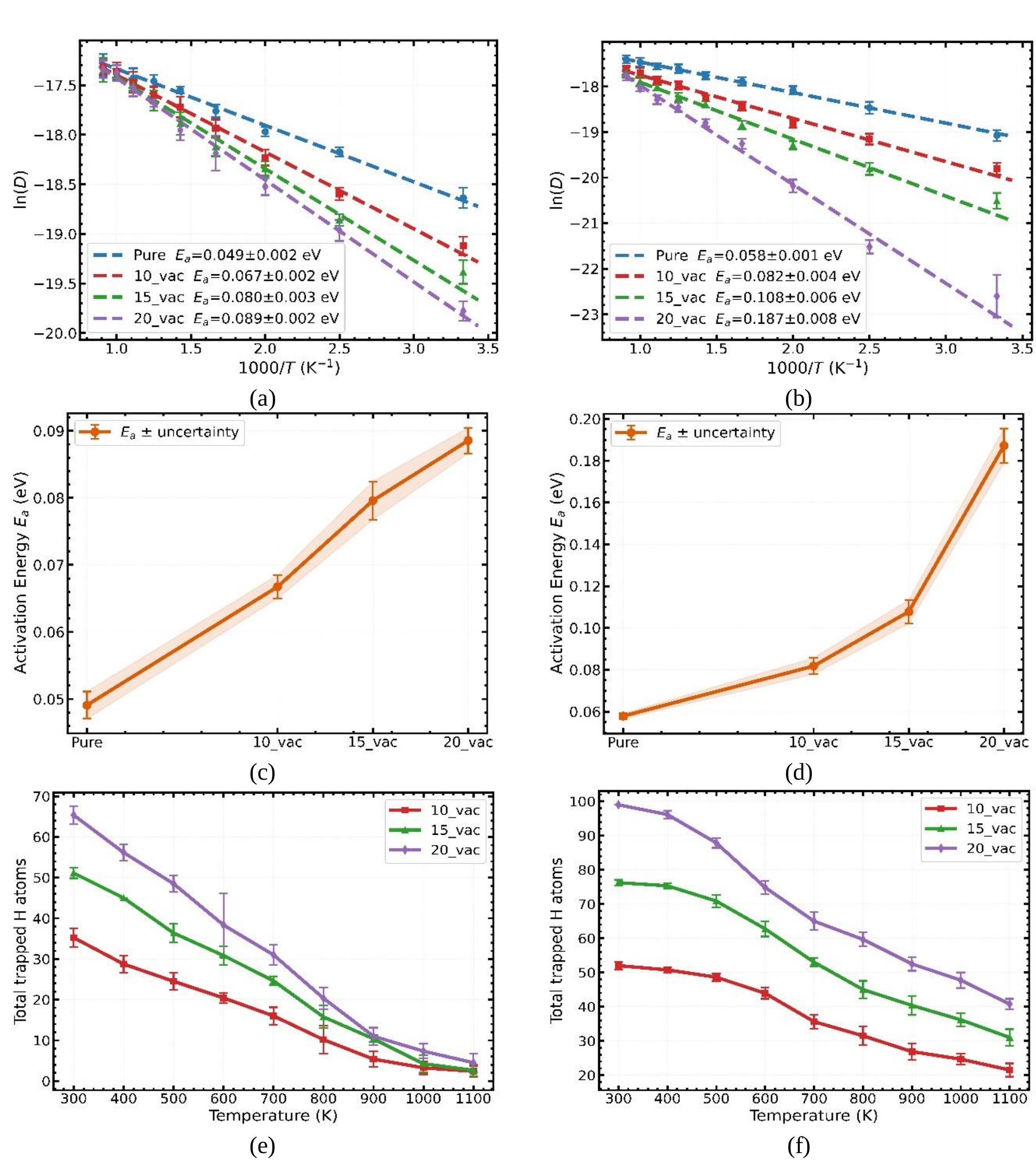

Figure 4, (a) and (b) illustrate the Arrhenius behavior of hydrogen diffusion in Fe and Cr. (c) and (d) represent the variation in activation energy with increasing vacancy concentration in Fe and Cr. (e) and (f) represent the total number of hydrogen atoms trapped at vacancy sites in Fe and Cr.

## 4. Conclusion

A multiscale approach combining density functional theory and kinetic Monte Carlo simulations was employed to investigate hydrogen–vacancy interactions and diffusion behavior in BCC Fe and Cr. Hydrogen is found to preferentially occupy tetrahedral sites and strongly interacts with vacancies, accommodating up to six hydrogen atoms per vacancy. The calculated binding energies and charge analysis indicate stronger hydrogen trapping in Cr compared to Fe. Climbing nudged elastic band calculations reveal detailed migration and detrapping pathways for multiple hydrogen occupancies. The detrapping energy generally decreases with increasing hydrogen content; however, in contrast to previous reports, the sixth hydrogen atom in Fe does not exhibit barrierless detrapping. In Cr, the trapping remains consistently strong, leading to higher energy barriers and enhanced retention of hydrogen. Kinetic Monte Carlo simulations demonstrate that vacancy concentration significantly reduces hydrogen mobility and increases the effective activation energy, with a more pronounced effect in Cr. The present results provide a systematic pathway level understanding of hydrogen trapping and detrapping in BCC metals and highlight important differences between Fe and Cr, offering insights relevant to hydrogen transport and embrittlement phenomena.

## Acknowledgments

I would like to sincerely thank Dr. Manan Dholakia and Dr. P. Ravindran for their valuable discussions. I also thank my colleagues, Meena and Prasanta, for their technical support. Finally, I thank the IGCAR Computer Division for providing HPC support.

.